\documentclass[a4paper,twocolumn]{esapub2005}
\usepackage{graphicx}
\usepackage{times}
\usepackage{xspace}
\usepackage{amsmath}
\usepackage{amssymb}
\usepackage{natbib}
\newcommand{\integral}{\textsl{INTEGRAL}\xspace}

\newcommand{\V}{V\,0332$+$53\xspace}
\newcommand{\fu}{4U\,1907$+$09}
\newcommand{\inte}{\textsl{INTEGRAL}}
\newcommand{\xte}{\textsl{RXTE}}

\newcommand{\kev}{\ensuremath{\text{keV}}}
\newcommand{\asm}{\textsl{ASM}}
\newcommand{\msun}{\ensuremath{\text{M}_{\odot}}}
\newcommand{\rsun}{\ensuremath{\text{R}_{\odot}}}
\newcommand{\lsun}{\ensuremath{\text{L}_{\odot}}}

\title{Accreting X-ray Pulsars observed with \integral}
\author[1,2]{Ingo Kreykenbohm}
\author[1]{Sonja Fritz}
\author[2]{Nami Mowlavi}
\author[3]{J\"orn Wilms}
\author[4]{Peter Kretschmar}
\author[1]{R\"udiger Staubert}
\author[1]{Andrea Santangelo}

\affil[1]{Institut f\"ur Astronomie und Astrophysik --
  Abt. Astronomie, Sand 1, 72076 T\"ubingen, Germany}
\affil[2]{Integral Science Data Centre, Chemin d'Ecogia 16, 1290
  Versoix, Switzerland}
\affil[3]{Dr. Remeis Sternwarte, Universit\"at Erlangen-N\"urnberg, Sternwartstr. 7, 96049 Bamberg, Germany}
\affil[4]{European Space Astronomy Centre (ESA), Madrid, Spain}

\begin{document}
\maketitle

\begin{abstract}
  In this paper, we review the observational properties of two
  accreting X-ray pulsars, the persistent source 4U\,1907$+$09 and the
  transient V\,0332$+$53.  Accreting X-ray pulsars are among the
  brightest sources in the X-ray sky and are frequently observed by
  Integral and other X-ray missions. Nevertheless they are still very
  enigmatic sources as fundamental questions on the X-ray production
  mechanism still remain largely unanswered. These questions are
  addressed by performing detailed temporal and spectral studies on
  several objects over a long time range.

  Of vital importance is the study of cyclotron lines as they provide
  the only direct link to the magnetic field of the pulsar. While some
  objects show cyclotron lines which are extremely stable with time
  and with pulse phase, in other objects the lines strongly depend on
  the pulse phase, the luminosity of the source, or both.

  Of special interest in any case are transient sources where it is
  possible to study the evolution of the cyclotron line and the source
  in general from the onset until the end of the outburst through a
  wide range of different luminosity states.

  We present an observational review of a transient and a persistent
  accreting X-ray pulsar observed with Integral.
\end{abstract}

\section{Introduction}

Accreting X-ray pulsars are among the most prominent sources in the
X-ray sky. Since accreting X-ray pulsars are usually young objects,
they are clustered along the galactic plane. Although already
discovered in the first days of X-ray astronomy, many fundamental
questions remain unanswered to date. The X-ray production mechanism is
only roughly understood, as a detailed understanding would require a
full magneto-hydrodynamic treatment, incorporating the presence of a
magnetic field of $\sim$$10^{12}$\,G, and a full handling of general
relativity. Especially the behavior and shape of cyclotron resonant
scattering features (CRSFs or cyclotron lines) are not very well
understood and require detailed studying.  But also only a very
general picture of the overall accretion geometry and the accretion
process itself exists. Analyzing the temporal behavior of accreting
neutron stars along with the spectral properties allows to study these
important questions. 

\integral is an excellent instrument to study these sources as
\integral frequently observes the galactic center and the galactic
plane and due to its large field of view. As a result, an enormous
amount of data on accreting X-ray pulsars is obtained.  In this paper
we present some recent results on the analysis of two accreting X-ray
pulsars: the detection of a complete torque reversal in \fu\
\citep{fritz06a} and the evolution of the transient \V over its recent
outburst and especially the behavior of the cyclotron line
\citep{mowlavi05a,tsygankov06a}.

\section{\fu}

The wind-accreting High Mass X-ray Binary system 
\fu\ \citep{giacconi71b} consists of a neutron star in an eccentric
($e=0.28$) 8.3753\,d orbit around its companion.
Based on a lower limit for the distance of 5\,kpc, the X-ray
luminosity is approximately
$2\,10^{36}\,\text{erg}\,\text{s}^{-1}$ \citep{zand97a}.
The stellar companion has been classified as
a O8--O9 Ia super-giant with an effective temperature of 30500\,K, a
radius of 26\,\rsun, a luminosity of $5\,10^{5}\,\lsun$, and a mass
loss rate of $7\,10^{-6}\,\msun\,\text{yr}^{-1}$ \citep{cox05a}.

\begin{figure}  
  \includegraphics[width=\columnwidth]{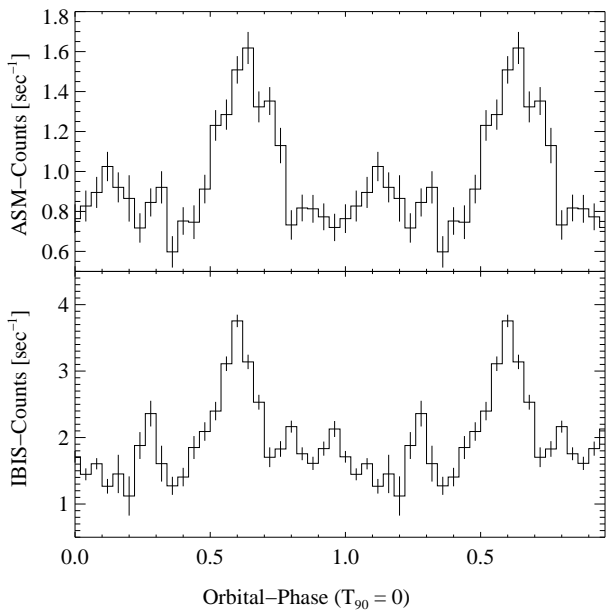}
  \caption{Light curve of \fu, folded with the orbital period of
    8.3753\,days \citep[from][]{fritz06a}. Upper Panel: 
Rossi X-ray Timing Explorer All Sky Monitor
    (\asm) 2--12\,keV. Lower Panel: IBIS 20--40\,keV light
    curve.}\label{fig:asm}
\end{figure}

Similar to other accreting neutron stars, the X-ray continuum of \fu\
can be described by a power-law spectrum with an exponential turnover
at $\sim$13\,keV \citep{mihara95a}.  The spectrum is modified by
strong photoelectric absorption with a column $N_\text{H}=1.5$--$5.7\,
10^{22}\,\text{cm}^{-2}$ \citep{cusumano98a} which is due to the dense
stellar wind and is therefore strongly variable over the orbit.  The
column density is maximal between the end of the primary X-ray flare
and the start of the secondary flare (Fig.~\ref{fig:asm}) of the
orbital light curve \citep{roberts01a}.  At higher energies, the
spectrum exhibits cyclotron resonant scattering features (CRSF) at
$\sim$19\,\kev\ and $\sim$40\,keV
\citep{makishima92a,mihara95a,cusumano98a}.

With a pulse period of $\sim440$\,s, \fu\ is a slowly rotating neutron
star \citep{makishima84a}.  The neutron star has exhibited a steady
linear spin down with an average of
$\dot{P}_\text{pulse}=+0.225\,\text{s}\,\text{yr}^{-1}$ from
$P_{\text{pulse}}=437.5$\,s in 1983 to 440.76\,s in 1998
\citep{zand98a}. Recently, a decrease in $\dot{P}_\text{pulse}$ to
$\sim$$0.115\,\text{s}\,\text{yr}^{-1}$, about half the long term
value was reported \citep{baykal06a}.

\subsection{Timing Analysis}
\label{sec:lc}
\begin{figure}  
  \includegraphics[width=\columnwidth]{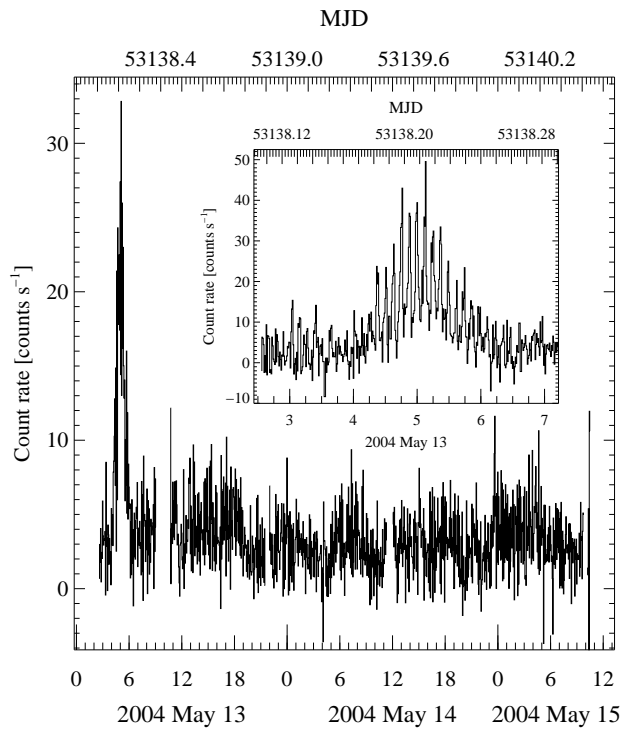}
  \caption{IBIS (ISGRI) 20--40\,\kev\ light curve of revolution 193
    \citep[from][]{fritz06a}. The bin
  time is 200\,s. The inset shows a close up of the flare, the bin
  time in this case is 40\,s. X-axis numbers are hours.}
   \label{fig:193lcinset}
\end{figure}

The X-ray light curve of \fu\ clearly shows pulsations with a
period of $\sim$441\,s. Additionally, \fu\ also exhibits flares with a
typical length of several hours \citep{makishima84a,zand98a}.
Fritz et al. \cite{fritz06a}
observed four such flares with count rate increases of at least
$5\sigma$ over its normal level (Fig.~\ref{fig:193lcinset}). Three of
them, in revolutions 187, 193, and 305, are associated with the main
peak in the orbital light curve (Fig.~\ref{fig:asm}) while the flare
observed in revolution 188 is linked to the secondary peak.

\begin{figure}
  \includegraphics[width=\columnwidth]{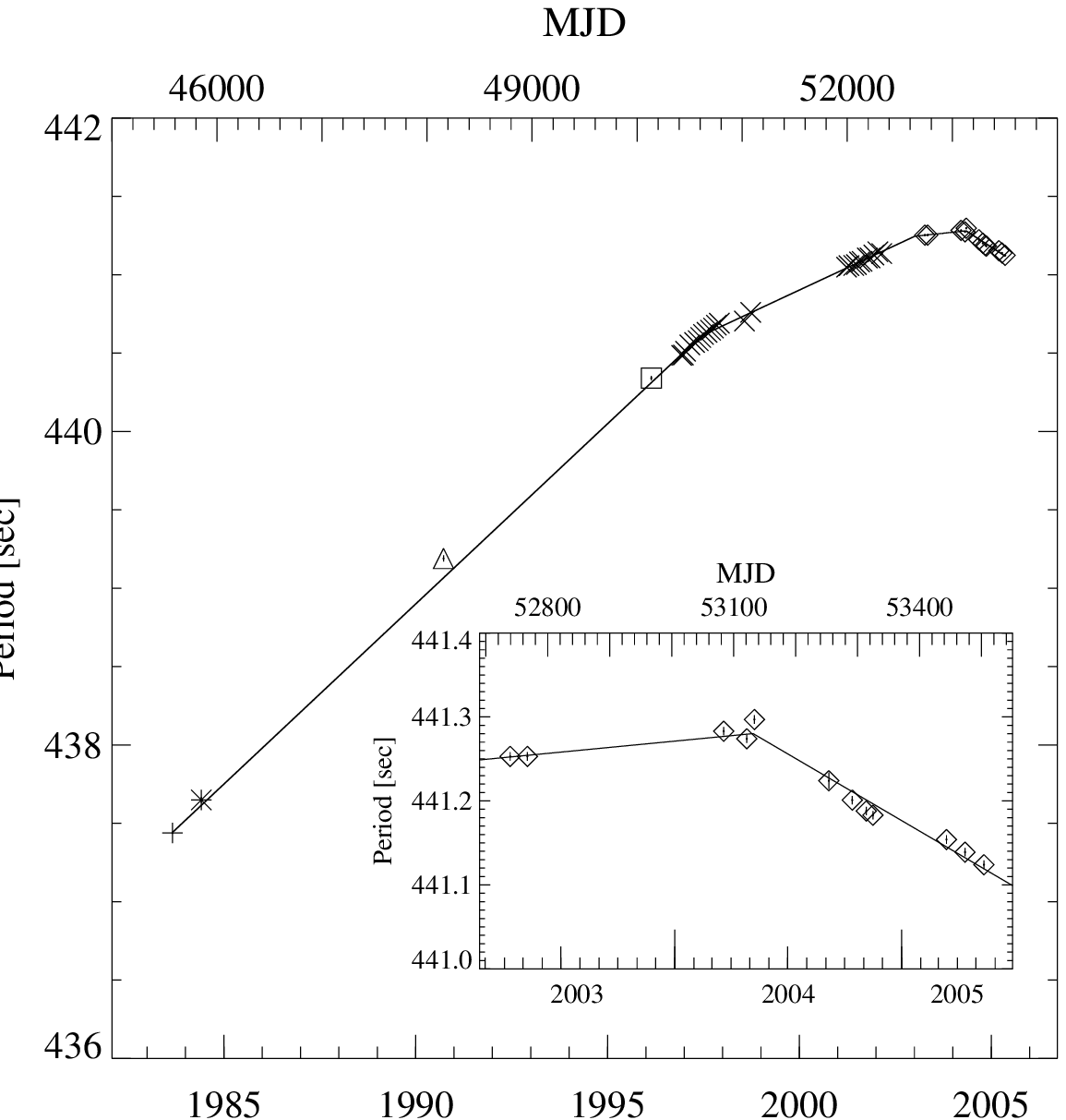}
 \caption{Evolution of the \fu\ pulse period over the last 20 years \citep[as
   shown by][]{fritz06a}. Diamonds
   indicate the results obtained in this work, other symbols
     indicate data from \cite[][plus-signs]{makishima84a},
     \cite[][asterixes]{cook87a}, \cite[][triangles]{mihara95a}, \cite[][squares]{zand98a}, and
     \cite[][crosses]{baykal01a,baykal06a}. Note the clear deviation
     from the long term trend leading to a complete spin reversal. The
   inset shows a closeup of the periods found by \inte\ \cite{fritz06a}.}
   \label{fig:perievo}
\end{figure}

\label{sec:pulsevol}

\fu\ showed a steady spin down rate of
$\dot{P}_\text{pulse}=+0.225\,\text{s}\,\text{yr}^{-1}$
\citep{zand98a}.  Baykal et al. \cite{baykal06a} using \xte-data
determined a spin down rate 0.5 times lower than the long term
value. Fritz et al. \cite{fritz06a} using \inte\ data, extend the
tracking of the pulse period evolution to the years 2003--2005.
Therefore these authors first determine the pulse period by first
using epoch folding \citep{leahy83a} to get a starting period and then
performing pulse phasing to get a higher
accuracy. Fig.~\ref{fig:perievo} shows the long-time history of the
period evolution based on all available data.  Note that the recent
periods obtained from\xte-data \cite{baykal06a} show a clear deviation
from the historic spin down trend of $\dot
P_{\text{pulse}}=+0.225\,\text{s}\,\text{yr}^{-1}$ from
\cite{zand97a}. The periods obtained by Fritz et
al. \cite{fritz06a} from \inte-data (shown as diamonds) not only
confirm this change of the trend, but show a complete trend reversal
from the historic long term spin down trend to a spin up trend from
MJD\,53131 onwards (see Fig.~\ref{fig:perievo}).

\begin{figure}  
  \includegraphics[width=\columnwidth,height=0.91\textheight]{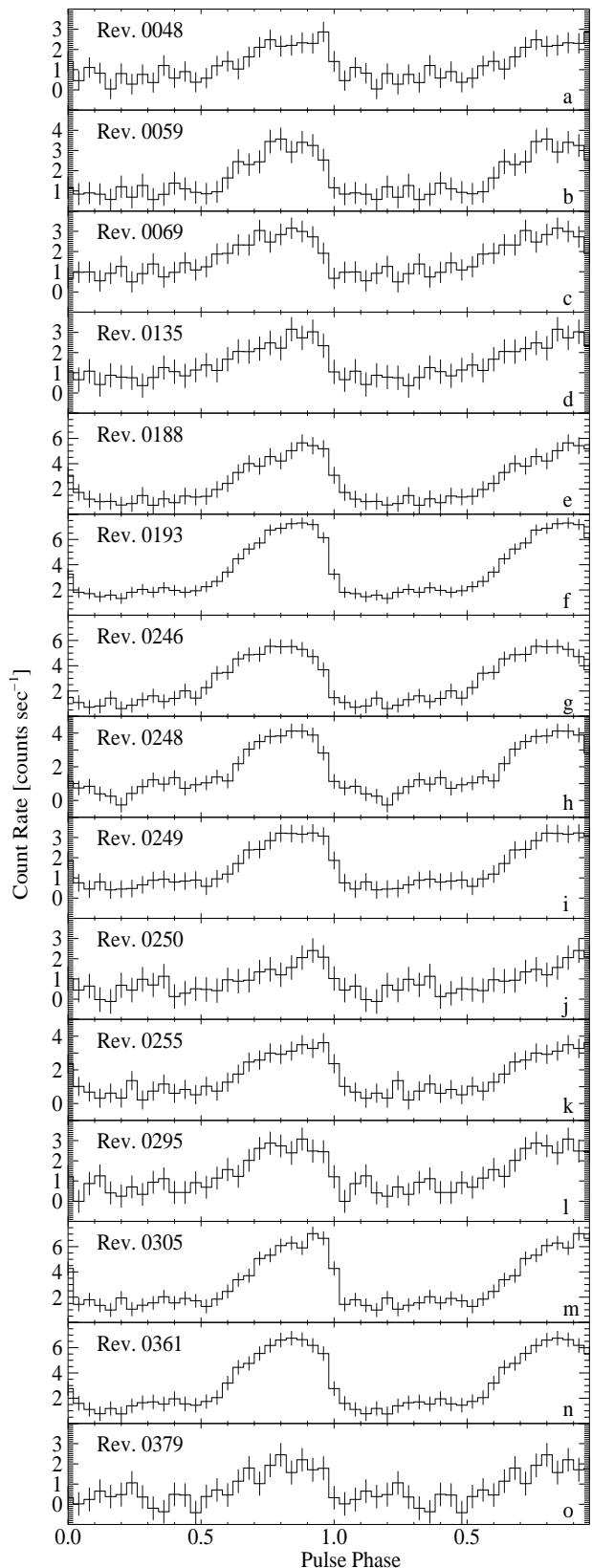}
  \caption{Pulse profiles for different revolutions in the
    20--40\,\kev\ band \citep[from][]{fritz06a}. Panels a--e show 
    pulse profiles obtained during the period of spin down, and panels f--o 
    show spin up pulse profiles. }
   \label{fig:pp_rev}
\end{figure}

Using the derived periods, Fritz et al. \cite{fritz06a} obtained pulse
profiles for all revolutions (Fig.~\ref{fig:pp_rev}) by folding the
light curves with the best respective periods. No change in the shape
of the pulse profile is obvious, especially no change in the shape
before, during, and after the reversal of $\dot{P}$ can be seen.

\subsection{Discussion}\label{sec:summary}

In the \inte\ data, a spin up phase of \fu\ is clearly detected after
20 years of spin down (Fig.~\ref{fig:perievo}) at a rate of
$\dot{P}_{\text{pulse}}=+0.225\,\text{s}\,\text{yr}^{-1}$, with the
available data being consistent with $\ddot{P}=0\,\text{s}^{-1}$. In
2002 and 2003 first indications for a decrease in the magnitude of
spin down were detected \citep{baykal06a}.  As shown in
Sect.~\ref{sec:pulsevol}, the new \inte\ data show a torque reversal
from $\sim$MJD\,53131 onwards, with the source now exhibiting a spin
up with a rate of
$\dot{P}_\text{pulse}=-0.158\,\text{s}\,\text{yr}^{-1}$.  So far, the
\inte\ results appear to be consistent with
$\ddot{P}=0\,\text{s}^{-1}$ (the $1\sigma$ upper limit for $\ddot{P}$
during the spin up is $-4\,10^{-5}\,\text{s}^{-1}\,\text{yr}^{-2}$).

The conventional interpretation of spin up in accreting X-ray pulsars
with a strongly magnetized, disk-accreting neutron star is that the
accretion disk is truncated by the magnetic field of the neutron star
\citep[see, e.g.,][and references
therein]{ghosh78a,ghosh79b,ghosh79c}. The transfer of angular momentum
from the accreted matter onto the neutron star will lead to a torque
onto the neutron star resulting in spin up.  The general expectation
of these models is that the Alfv\'en radius $r_\text{m}$, where the
accretion flow couples to the magnetic field, is assumed to be close
to the co-rotation radius, $r_\text{co}$.

This simple model, however, is unable to explain the long episodes of
constant $\dot P$.  For \fu\ the magnetospheric radius inferred from
the cyclotron line measurements is $r_\text{m}\sim 2400$\,km, while
$r_\text{co}\sim12000$\,km \citep{zand98a}. Moving $r_\text{m}$ out to
$r_\text{co}$ would require a magnetic field of
$\sim$$10^{14}\,\text{G}$ as for a magnetar, which is two orders of
magnitude larger than the magnetic field deduced from the observed
cyclotron lines.

Fritz et al. \cite{fritz06a}, however, then discuss that the a model
presented recently by Perna et al. \cite{perna06a} is able to explain
these observations. This model is based on the fact that X-ray pulsars
are oblique rotators, i.e., the neutron star's magnetic field is
tilted by an angle $\chi$ with respect to the axis of rotation of the
neutron star. Assuming the accretion disk is situated in the neutron
star's equatorial plane, for a ring of matter in the accretion disk,
the magnetic field strength then depends on the azimuthal angle and
thus the boundary of the magnetosphere is asymmetric. Such a
configuration can lead to regions in the disk where the propeller
effect is locally at work, while accretion from other regions is
possible. This then results in a nonlinear dependence of the accretion
luminosity from the $\dot{M}$ through the outer parts of the disk.
Perna et al. \cite{perna06a} point out that a nontrivial consequence
is that for values of $\chi$ greater than a critical value,
$\chi_\text{crit}$, limit cycles are present, where cyclic torque
reversal episodes are possible without a change in $\dot{M}$. For
typical parameters. $\chi_\text{crit}$ is between $\sim$$25^\circ$ and
$\sim$$45^\circ$.  Perna et al. \cite{perna06a} and Fritz et
al. \cite{fritz06a} stress that the existence of the limit cycles
depends only on $\chi$ and on the pulsar's polar magnetic field, $B$,
and no variation of external parameters is required to trigger torque
reversals, while the torque reversal of \fu\ with no associated drop
in luminosity and no change in the shape of the pulse profile is
difficult to explain in conventional models.

\section{\V}

The recurrent X-ray transient V\,0332+53 experienced an X-ray outburst
from December 2004 to February 2005
\citep{kreykenbohm05a,pottschmidt05a}.  The outburst was predicted
almost one year before the actual outburst \cite{goranskij04a} due to
the optical brightening of the optical companion BQ~Cam, a O8-O9e
star.

The observation of this outburst of V\,0332+53 by \xte\ revealed the
presence of three cyclotron lines
\citep{coburn05a,kreykenbohm05a,pottschmidt05a}. The fundamental CRSF
has an energy of 28\,keV, and the first and second harmonics are at
50\,keV and 71\,keV, respectively. This corresponds to a magnetic
field strength of $2.7 \times 10^{12}$\,G
\citep{kreykenbohm05a,pottschmidt05a}.  The spectrum is otherwise
described by a power law modified by a high-energy cut-off at higher
energies.

\subsection{Fluxes}
\label{Sect:flux}

\begin{figure}
    \centering
    \includegraphics[width=\columnwidth]{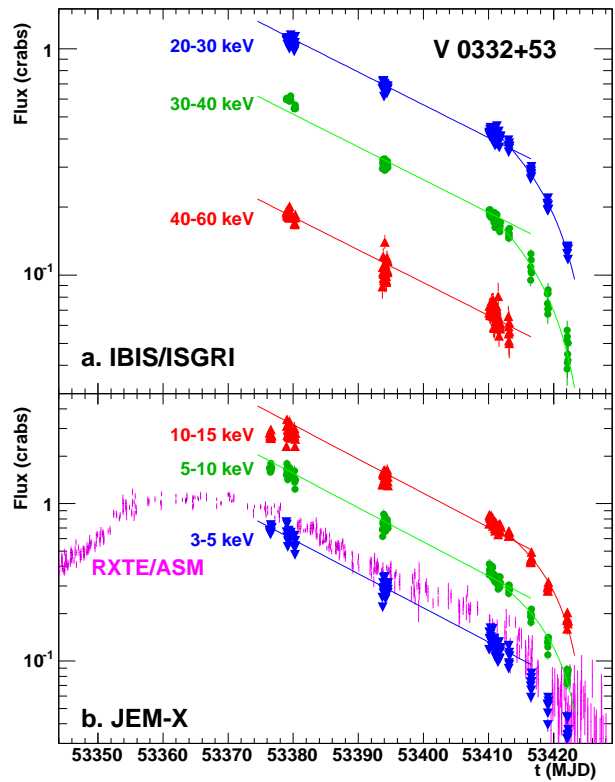}
    \caption{\textbf{a} Light curve of \V over the decay of the
      outburst in different energy bands in Crab units \citep[as shown
      by][] {mowlavi05a}.  The solid curves are fits to the
      exponential (from MJD=53376 to 53412 with $\tau$=30 days) and
      linear (from MJD=53412 to 53422) \textbf{b} Same as \textbf{a},
      but for JEM-X fluxes and with an exponential decay of $\tau$=20
      days.  The 2--12\,keV \xte/\asm\ light curve is also shown,
      using a normalization factor of 77 cps for 1 Crab.  }
    \label{Fig:flux}
\end{figure}

\begin{figure}
\includegraphics[width=\columnwidth]{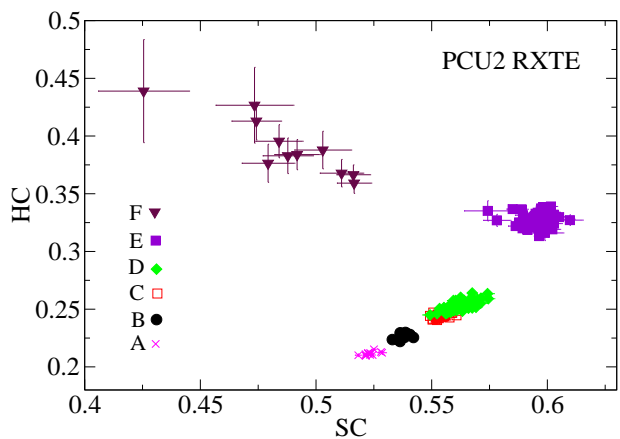}
\caption{Color-Color Diagram of the outburst of \V 
  \citep[from][]{reig06a} using \xte-data. The
  X-ray colors are defined as soft color SC=7.5-10 keV/4-7.5 keV
  and hard color HC=15-30 keV/10-15 keV.  }
\label{color}
\end{figure}

The light curve of \V in ISGRI and JEM-X during the outburst is shown
in Fig.~\ref{Fig:flux}. Mowlavi et al. \cite{mowlavi05a} describe the
decline phase by an exponential decay up to MJD$\simeq$53412, followed
by a linear decrease. The exponential decay is less rapid at the ISGRI
energies than at the JEM-X energies.  From a simple estimation, the
exponential decay time is about 30 days above 20\,keV (solid curves in
Fig.~\ref{Fig:flux}a) and about 20 days below 15\,keV (solid curves in
Fig.~\ref{Fig:flux}b).  The exponential decay is also confirmed by the
2--12\,keV light curve as observed by the All Sky Monitor (\asm) on
board of \xte\ (see Fig.~\ref{Fig:flux}b).

Reig et al. \cite{reig06a} discovered a very interesting behavior of
the color-color diagram over the outburst (see Fig.~\ref{color}). They
derived background-subtracted light curves corresponding to the energy
ranges $c_1=4-7.5$ keV, $c_2=7.5-10$ keV, $c_3=10-15$ keV and $c_4=
15-30$ keV and then defined the soft color (SC) as the ratio $c_2/c_1$
and the hard color (HC) as the ratio between $c_4/c_3$.

These authors observed that the source was in a soft state during the
peak of the outburst and as the count rate decreased the source moved
up and right in the color color diagram, i.e. it became harder. At the
end of the outburst (regions E and F) the source moved again to the
left, i.e., lower values of SC, while HC remained mostly
constant. Therefore two spectral states or branches can be
distinguished: a soft state when the source is bright and a hard state
when the source gets dimmer.

\begin{figure}
    \centering
    \includegraphics[width=\columnwidth]{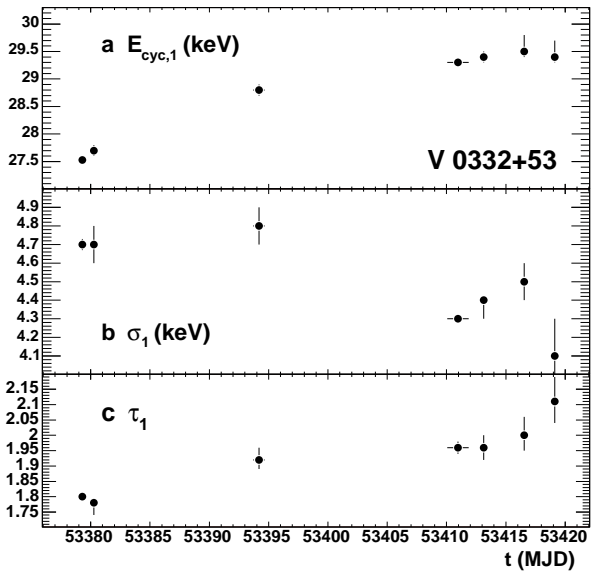}
    \caption{Evolution of the parameters of the fundamental cyclotron
      line \citep[from][]{mowlavi05a}. \textbf{a} line energy,
      \textbf{b} width, and \textbf{c} depth. The vertical bars
      represent 90\% uncertainty, while the horizontal bars indicate
      the duration of each observation.  }
    \label{Fig:spefit}
\end{figure}

The pattern in the color color diagram of \V resembles that of low
mass X-ray binary Z source \citep[see e.g][]{vanderklis95a}. In this
picture, the soft branch would correspond to the normal branch and the
hard branch to the horizontal branch in a Z source, however, as Reig
et al. \cite{reig06a} point out, the flaring branch observed in Z
source is missing in V\,0332+53.  During the decline of the outburst,
the source moves without jumps through the Z shape in the color color
diagram, taking about 80 days from the softest to hardest point
\citep{reig06a}.

\subsection{Spectrum}
\label{Sect:spectrum}

The spectra of accreting X-ray pulsars are a superposition of
contributions from the hotspots at both magnetic poles and the lower
parts of the accretion columns. Since no self consistent model is
available, various empirical models have to be used \citep{kreykenbohm99a}

\begin{figure}
    \centering
    \includegraphics[width=\columnwidth,height=0.7\textheight]{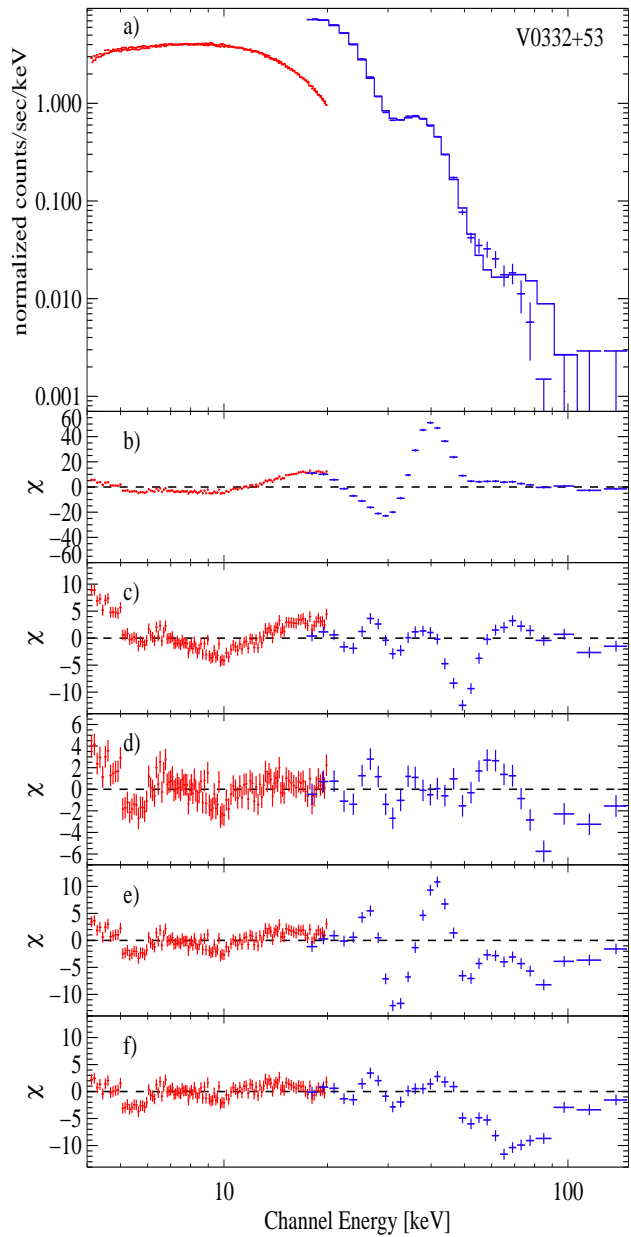}
    \caption{\textbf{a} Spectrum and folded model for JEM-X (left) and
      ISGRI (right) using data of revolution 284
      \citep[from][]{mowlavi05a}; \textbf{b} residuals of the model
      with $\Gamma$=$-0.41$ and $E_{\rm fold}$=7.4\,keV, and without
      any cyclotron line modeled; \textbf{c} one Gaussian is included
      at 29.3\,keV; \textbf{d} a second Gaussian is added at 51\,keV
      to model the first harmonic; \textbf{e} as before, but with the
      energy of the first line fixed at 27.5\,keV, the value found in
      revolution 273; \textbf{f} same as \textbf{d}, but with $\Gamma$
      fixed to the value of $-0.17$ found in revolution 273. The
      strong residuals in the two bottom panels show that the
      parameters of the fundamental cyclotron line do indeed change. }
    \label{Fig:residuals}
\end{figure}

Mowlavi et al. \cite{mowlavi05a} use the simple ``cutoffpl''
model to fit the combined ISGRI and JEM-X data, to which two cyclotron
lines with a Gaussian optical depth profile \citep{coburn02a} have
been added.  A multiplicative constant is applied to allow for the
different normalizations of the two instruments.  Furthermore a
systematic error of 2\% is applied to account for the uncertainties in
the response matrices of JEM-X and ISGRI.  No photoelectric
absorption is required at low energies, compatible with the fact that
the JEM-X data are analyzed only above 4\,keV.  The same model is used
for all observations to keep consistency between the parameters from
one observation to the next.

The cyclotron line parameters for the fundamental line
\cite{mowlavi05a} are displayed in Fig.~\ref{Fig:spefit} as a function
of time. The spectrum with the fits and the residuals for a model with
no, one, two, and three cyclotron lines using the data of revolution
284 is shown in Fig.~\ref{Fig:residuals}.  The energy of the
fundamental cyclotron line increases by 7\% from the first to the last
\textsl{INTEGRAL} observations, and the depth of the line increases by
16\% while at the same time the width decreases by 12\%. These changes
are significant; according to Mowlavi et al. \cite{mowlavi05a}, the
attempt to fit the spectrum in revolution 284 with a line at 27.5\,keV
results in strong residuals around the line
(Fig.~\ref{Fig:residuals}e), but also variations of continuum can be
ruled out to explain the changes in the line parameters
(Fig.~\ref{Fig:residuals}f).

\begin{figure}
\centerline{\includegraphics[width=\columnwidth]{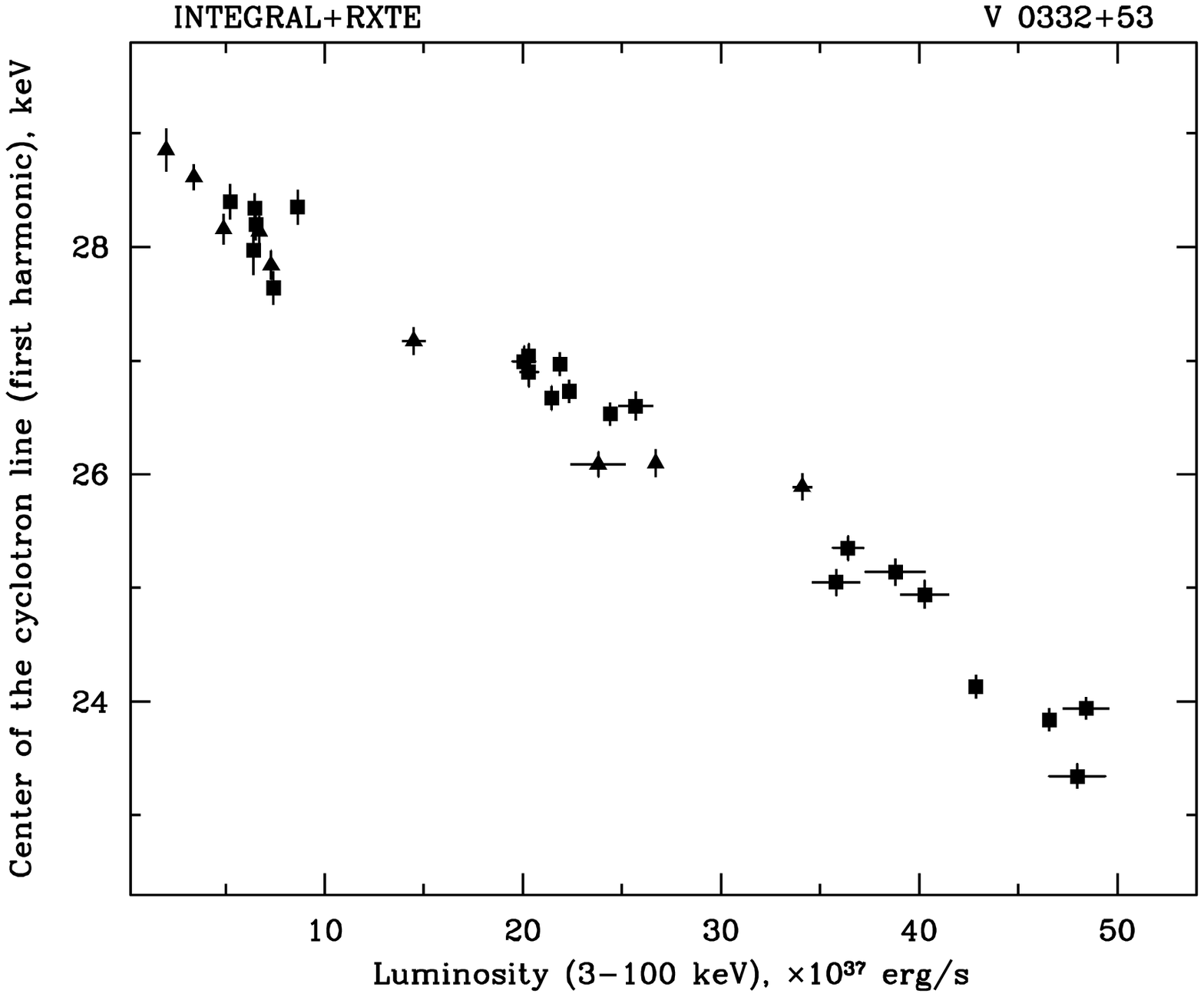}}
\caption{Dependence of the energy of the energy of the fundamental
  cyclotron line of \V \citep[from][]{tsygankov06a}. Triangles
  represent results from \integral, and squares from \xte.}
\label{Fig:tsygankov}
\end{figure}

Tsygankov et al. \cite{tsygankov06a} then went one step further and
instead plotting the energy of the fundamental cyclotron line versus
time, these authors show the dependence of the cyclotron energy on the
luminosity of the source, thus combining Figs.~\ref{Fig:spefit} and
\ref{Fig:flux} into Fig.~\ref{Fig:tsygankov}.  A linear fit to this
dependence results in the following relation: $$E_\text{cyc,1}=-0.10
L_{37} + 28.97 \text{keV},$$ where $L_{37}$ is the luminosity in units of
$10^{37}$\,erg s$^{-1}$. Assuming that for low luminosities the
emission comes practically from the neutron star surface, Tsygankov et
al. \cite{tsygankov06a} estimate the magnetic field on the surface of
the neutron star for canonical values for the radius and the mass of
the neutron star to $B = 3.0 \ 10^{12}$\,G.

\subsection{Discussion}
\label{Sect:discussion}

V\,0332+53 is a highly magnetized pulsar with a magnetic field of about
$3.0\times 10^{12}$\,G
\citep{kreykenbohm05a,pottschmidt05a,tsygankov06a}, as derived from
the energy of the fundamental cyclotron line.  The X-ray outbursts
from this system are observed with a time recurrence of about 10
years.  They are believed to be triggered by a massive mass ejection
from the optical companion star, resulting in the optical brightening
of the system in 2004 \citep{goranskij04a}.  The ejected mass feeds an
accretion disk around the pulsar, from which it is funneled by the
magnetic field towards the polar caps and falls on the surface of the
neutron star.

The analysis of the 2005 outburst provides new insights towards a
better understanding of V\,0332+53 and similar transient sources
within this basic picture.  The X-ray light curve
(Sect.~\ref{Sect:flux}) displays a characteristic exponential decay
with a time scale of 20 to 30 days, followed by a linear
decrease. This remarkable change occurs approximately at the
same moment when the trend in the color color diagram changes. This may
provide some information on the accretion and especially the geometry.

As discussed by Mowlavi et al. \cite{mowlavi05a}, an exponential decay
of the flux is observed also in dwarf novae and in soft X-ray
transients on time scales from a few days for the former to several
weeks for the latter \citep{lasota96a,tanaka96a}. King \& Ritter
\cite{king98a} show how illuminated disks in low-mass X-ray binary
stars can produce such an exponential decay on the time scales
observed for the soft X-ray transients. Subsequently, numerical
calculations by Dubus et al. \cite{dubus99a} showed, that indeed
irradiated accretion disks produce an exponential decay of the X-ray
luminosity.

As Mowlavi et al. \cite{mowlavi05a} point out, it is tempting to make
the parallel with the results observed in \V.  The exponentially
decreasing flux followed by a linear decrease is very similar to what
is observed in disk-illuminated systems. Such a behavior as observed
by Mowlavi et al. \cite{mowlavi05a} has never before been observed in a
high mass accreting X-ray binary pulsar system. The X-ray flux is
proportional to the mass accretion rate, and therefore the accretion
rate would then also be proportional to the mass of the disk. An
unknown change in the disk would then trigger the switch to the linear
decay phase. Furthermore, Mowlavi et al. \cite{mowlavi05a} suggest
that the two different decay times observed in \V are due to the
presence of at least two regions contributing to the X-ray continuum:
one time scale governs the rate of mass flow onto the neutron star
through the disk, leading to the exponential decay followed by a
linear decrease (see above); another time scale governs the spectral
changes of the emission that may come from different heights within
the accretion column \citep{mowlavi05a}.

The X-ray spectrum (Sect.~\ref{Sect:spectrum}) is well fitted by a
cutoff power law modified by two cyclotron lines.  The energy of the
fundamental cyclotron line is shown to increase by 20\% from the peak
of the outburst as measured by \xte\ until the end of the outburst.
At the same time it became narrower and deeper.  This information
would provide some insight into the resonance scattering region near
the polar caps.  The evolution of cyclotron line parameters with time
has already been reported in the literature for several binary systems
\citep{mihara95a}.  These authors attribute the change of the line
energy to a change of the height of the scattering region above the
neutron star, resulting in a height change $\sim$300\,m for V~0332+53.
Kreykenbohm et al. \cite{kreykenbohm04a} report a variation of the
cyclotron energy of GX\,301-2 with the pulse phase of the pulsar, and
attribute the change to different viewing angles of the accretion
column where the line originates.

For \V Mowlavi et al. \cite{mowlavi05a} interpret the increase of the
energy as a variation of the scattering region characteristics.  They
assume that the region where the cyclotron resonance scattering takes
place is extended. The region would then cover a range of magnetic
field strengths resulting in a superposition of several narrow
cyclotron lines, each originating at a specific height and therefore
in a specific magnetic field strength, thus giving rise to a broad
line.  As the accretion rate and hence the flux decreases with time,
the cyclotron resonance region shrinks and moves closer to the neutron
star. At the end of the outburst only the region closest to the
neutron star surface, where the magnetic field is strongest, would
contribute to the formation of cyclotron lines. The resulting lines
can therefore be expected to be observed at higher energies and
being much narrower as is indeed observed in \V. Mowlavi et
al. \cite{mowlavi05a} estimate the total movement of the cyclotron
line formation region to be of the order of 500\,m, similar to
4U\,0115$+$63 where a change of $\sim$1\,km was observed
\citep{mihara98a}.

\section*{Acknowledgments}
  We acknowledge the support of the Deutsches Zentrum f\"ur Luft- und
  Raumfahrt under grant numbers 50OR0302, 50OG9601, and 50OG0501, and
  by National Aeronautics and Space Administration grant
  INTEG04-0000-0010.

\bibliography{xpuls,velax1,div_xpuls,misc,roentgen,books,foreign}

\end{document}